\documentclass[12pt,a4paper]{article}
\usepackage{t1enc,epsfig}
\usepackage[english]{babel}

\begin{document}
\def\vA{{\bf A}}
\def\vf{{\bf f}}
\def\vk{{\bf k}}
\def\vJ{{\bf J}}
\def\vn{{\bf n}}
\def\vp{{\bf p}}
\def\vq{{\bf q}}
\def\vr{{\bf r}}
\def\vx{{\bf x}}
\def\vv{{\bf v}}
\newcommand{\ba}{\begin{eqnarray}}
\newcommand{\ea}{\end{eqnarray}}
\newcommand{\be}{\begin{equation}}
\newcommand{\ee}{\end{equation}}

   \title{ANOMALOUS TRANSPORT PROCESSES IN CHEMICALLY ACTIVE RANDOM
ENVIRONMENT}

   \author{J. Honkonen\thanks{Juha.Honkonen@helsinki.fi}\\
Department of Physical Sciences,
University of Helsinki,\\ and\\ National Defence College, Helsinki, Finland}


\date{\today}
\maketitle

\begin{abstract}
The effect of random velocity field on the kinetics of single-species
and two-species annihilation reactions is analysed near two dimensions
in the framework of the field-theoretic renormalisation group.
Fluctuations of particle density are modeled within the approach of Doi.
The random incompressible velocity field is generated by stochastically
forced Navier-Stokes equation in which thermal fluctuations - relevant
below two dimensions - are taken into account.
\end{abstract}


   \section{Introduction}

The effect of density fluctuations on the asymptotics of
reaction rates in low dimensionalities has attracted considerable
attention recently~\cite{Lee94,Lee95}.
Reaction rates may also be affected by fluctuations of an advective velocity field.
Most work in this direction has been carried out for the
case of quenched random drift~\cite{Oerding96,Deem98,Richardson99}.
Recently, the asymptotic behaviour of the unimolecular reaction
$A+A\to\emptyset$ in a dynamically generated random drift has
been analysed with the aid of field-theoretic renormalisation group (RG) \cite{Hnatic00}.
In my report I shall describe a similar approach to the bimolecular reaction
$A+B\to\emptyset$.

Classical rate equations for densities $n_A$, $n_B$ with the homogeneous initial
condition $n_A(0)=n_{A0}$, $n_B(0)=n_{B0}$
\begin{equation}
\label{classicrate}
{d n_A\over d t}=
-K_{0}n_A n_B\,,\qquad
{d n_B\over d t}=
-K_{0}n_An_B
\end{equation}
yield the ''normal'' decay laws. Let, for definiteness, $n_{A0}<n_{B0}$, then
\begin{equation}
\label{AB}
n_A(t){\scriptstyle\,\,\atop{\displaystyle\sim\atop \scriptstyle t\to\infty}}
(n_{B0}-n_{A0}){n_{A0}\over n_{B0}}\, {\rm e}^{-K_0(n_{B0}-n_{A0})t}
\,,\quad
n_B(t){\scriptstyle\,\,\atop{\displaystyle\sim\atop \scriptstyle t\to\infty}}(n_{B0}-n_{A0})\,.
\end{equation}
\medskip
In case of equal initial densities  $n_{A0}=n_{B0}=n_0$ a powerlike decay takes over:
\begin{equation}
\label{nn}
n_A(t)=n_B(t)\,{\scriptstyle\,\,\atop{\displaystyle\sim\atop \scriptstyle t\to\infty}}\,{1\over K_0t}\,.
\end{equation}
A heuristic account of initial-density fluctuations amplified by
diffusion together with numerical simulations~\cite{Ovchinnikov78,Toussaint83}
yield anomalous decay at $d<4$, e.g. for $n_{A0}=n_{B0}=n_0$:
\be
\label{d/4}
n_A(t)
\,{\scriptstyle\,\,\atop{\displaystyle\sim\atop \scriptstyle t\to\infty}}\,
{cn_0^{1/2}\over (Dt)^{d/4}}\,,
\ee
which is slower than the "classical" decay rate (\ref{nn}).

\section{Second quantization for reactions}

For a systematic analysis of the effect of density fluctuations
it is convenient to use a field-theoretic approach.
There are two possibilities available for this. First, the more
widely known Martin-Siggia-Rose (MSR)
approach~\cite{Martin73}, which corresponds to the solution of a Langevin-type stochastic
equation to describe fluctuations. Physically, this approach is well-suited
to situations in which fluctuations are of "external" origin, e.g.
small-scale thermal fluctuations from the point of view of
macroscopic (hydrodynamic) scale physics, or fluctuations caused
by some external random source. Second, the several times
reinvented Doi approach~\cite{Doi76}, in which the randomness is described by
a probability distribution functional (PDF) or, if you like,
an infinite set of probability distribution functions on a lattice,
with the subsequent set of master equations instead of Langevin
equations. This approach is better suited to cases, in which
fluctuations of intrinsic origin are dealt with. This is the
case, for instance, for density fluctuations due to randomness
in the (chemical) reaction process itself.

Therefore, I will use here the Doi approach, which allows for a
"mesoscopic" analysis of density fluctuations.
To calculate expectation values with the probability
distribution functional (PDF) $P(\{n_A({\bf x})\};\{n_B({\bf x})\},t)$
for the particle densities $n_A({\bf x})$ and $n_B({\bf x})$,
the formal solution of the set of master equations for the PDF
may be expressed in a functional form with the aid of
bosonic field operators
\cite{Doi76,Zeldovich78} with the commutation relations
\[
[\psi_A(\vx),\psi^+_A(\vx')]=\delta(\vx-\vx')\,,\quad
[\psi_A(\vx),\psi_A(\vx')]=[\psi^+_A(\vx),\psi^+_A(\vx')]=0\,,
\]
and a similar set for the $B$ particles. In a fairly standard-looking second-quantization
setting
the average of an observable $O$ may be written as a vacuum expactation value
\ba
\label{O}
\langle O(t)\rangle &=&\sum\limits_{\{n_i({\bf x})\}}O[\{n_A\},\{n_B\}]
P(\{n_A)\};\{n_B\},t)\nonumber\\
&=&\phantom{\Bigl(}\!\!\!\langle 0\vert O[(\psi^+_A+1)\psi_A,(\psi^+_B+1)\psi_B]\,
{\rm e}^{-\hat{H}'t}\\
&\times& {\rm e}^{\int\! d\vx\,(n_{A0} \psi^+_A
+n_{B0}\psi^+_B-r_0\sqrt{n_{A0} n_{B0}}\psi^+_A\psi^+_B)}\vert 0\rangle\,,\nonumber
\ea
with the kinetic operator
\ba
\label{H'}
\hat{H}'&=&\int\! d\vx\, \Biggl\{\psi^+_A\nabla (\vv\psi_A)+\psi^+_B\nabla (\vv\psi_B)
-D_{A0}\psi^+_A\nabla^2\psi_A-D_{B0}\psi^+_B\nabla^2\psi_B\nonumber\\
&+&
K_{0}\left(\psi^+_A+\psi^+_B+\psi^+_A\psi^+_B\right)\psi_A\psi_B
\Biggr\}\,.
\ea
The last exponential in (\ref{O}) corresponds to the initial PDF. A customary choice
is the Poisson distribution for the local particle number~\cite{Lee94,Lee95}.
However, in view of the hostile nature of interaction I have allowed for
negative initial correlations by using in (\ref{O})
a {\em bivariate} Poisson distribution~\cite{Feller71}. This choice
leads to the following expressions for the low-order moments
of initial densities:
\ba
\label{initial}
\setlength{\arraycolsep}{2pt}
\hspace{-3mm}\overline{n_A(\vx,0)}=n_{A0}\,,\ \
\overline{n_B(\vx,0)}=n_{B0}\,,\ \
\overline{\Delta n_A(\vx,0)\Delta n_B(\vx',0)}=-
r_0\sqrt{n_{A0}n_{B0}}\,\delta_{\vx,\vx'},\nonumber\\
\hspace{-7mm}
\overline{\Delta n_A(\vx,0)\Delta n_A(\vx',0)}=n_{A0}\,\delta_{\vx,\vx'}\,,\ \
\overline{\Delta n_B(\vx,0)\Delta n_B(\vx',0)}=n_{B0}\,\delta_{\vx,\vx'}.
\ea
Physically this corresponds to
thermal fluctuations with anticorrelations in initial reactant densities.

\section{Dynamic action for the advection-diffusion-controlled reaction $A+B\to\emptyset$}

Construction of
perturbation theory through the $T$ exponent for the
evolution operator
\[
U(t,t_0)={\rm e}^{\hat{H}'_0t}{\rm e}^{-\hat{H}'(t-t_0)}{\rm e}^{-\hat{H}'_0t_0}
=T\,{\rm e}^{-\int_{t_0}^t\!\hat{H}_I'dt}
\]
allows to write the expectation value (\ref{O}) as the following functional integral:
\be
\label{F1}
\langle O(t)\rangle=\int\!{\cal D}[\psi]\, O_N(1,\psi_A,1,\psi_B)
{\rm e}^{S_0+S_1}\,,
\ee
where $O_N$ is the normal symbol of the operator $O$
\[
O[\psi^+_A\psi_A,\psi^+_B\psi_B]=
N[O_N(\psi^+_A,\psi_A,\psi^+_B,\psi_B)]\,,
\]
$S_1$ is the dynamic action~\cite{Lee95}
\ba
\label{S1}
S_1&=&-\int\! d\vx dt\,\Bigl\{\psi^+_A\partial_t\psi_A+\psi^+_B\partial_t\psi_B+\psi^+_A\nabla (\vv\psi_A)
+\psi^+_B\nabla (\vv\psi_B)\nonumber\\
&-&D_{A0}\psi^+_A\nabla^2\psi_A-D_{B0}\psi^+_B\nabla^2\psi_B
+
K_{0}\left(\psi^+_A+\psi^+_B+\psi^+_A\psi^+_B\right)\psi_A\psi_B
\Bigr\}
\ea
and $S_0$ contains terms brought about by the initial
bivariate Poisson distribution
\[
S_0=\int\! d\vx\,\Bigl[ n_{A0}\psi^+_A+ n_{B0}\psi^+_B
-r_0\sqrt{n_{A0} n_{B0}}\psi^+_A\psi^+_B\Bigr]\,.
\]
Schwinger equations with respect to $\psi^+_A$, $\psi^+_B$
\ba
\left\langle{\partial_t\psi_A}
+\nabla (\vv\psi_A)-D_{A0} \nabla^2\psi_A
+K_{0}\psi_A\psi_B\right\rangle\!=n_{A0}\delta(t)\,,\\
\left\langle {\partial_t\psi_B}
+\nabla (\vv\psi_B)-D_{B0} \nabla^2\psi_B
+K_{0}\psi_A\psi_B\right\rangle\!=n_{B0}\delta(t)\,,
\ea
in the mean-field approximation
yield the classical rate equations (\ref{classicrate})
for the homegeneous average densities
$
\langle n_A(t)\rangle= \langle \psi_A(t)\rangle
$, $\langle n_B(t)\rangle= \langle \psi_B(t)\rangle
$.
It should be borne in mind, however, that the second and higher order
moments of the
fields $\psi_A$ and $\psi_B$
are not equal to the corresponding moments of the densities.
For instance, the pair correlations of $A$ particles
are given by
$
\langle n_A(t,{\bf x})
n_A(t,{\bf x}')\rangle=
\langle [\psi_A(t,{\bf x})\psi_A(t,{\bf x}')+\delta({\bf x}-{\bf x}')\psi_A(t,{\bf x})]\rangle
$.

To describe fluctuations of the drift field $\vv$ in (\ref{S1}) I use
random velocity field
generated by the transverse
stochastic Navier-Stokes equation
\be
\label{NS}
 \partial_t {\bf v} +
 P({\bf v} \cdot {\nabla} ) {\bf v} -
 \nu_0 \nabla^2 {\bf v}={\bf f}^v
\ee
with the incompressibility conditions:
 ${ \nabla}\cdot {\bf v}   = 0$,
 ${ \nabla}\cdot {\bf f}^v = 0$.
For the random force the Gaussian distribution
with zero mean
and the correlation function
\be
\label{corrf}
 \langle
  \, f^v_m (\vx_1,t_1) f^v_n (\vx_2,t_2)
  \, \rangle =
\delta(t_1-t_2)
 \int \frac{d{\bf k}}{(2\pi)^d}\,
 P_{mn}({\bf k})d_f(k)
 {\rm e}^{i {\bf k}   \cdot( {\bf x}_1 -{\bf x}_2)}
\ee
is assumed.
In (\ref{corrf}) $P_{mn}({\bf k})= \delta_{mn}- k_m k_n / k^2$
is the transverse projection operator in the wave-vector space, and $d_f(k)$
is a function of the wave number $k$ and the parameters of energy pumping,
which is used to produce stationary random drift.  The kernel function
is often chosen in the nonlocal form
\begin{equation}
\label{df}
d_f(k)=g_{10}\nu_0^3k^{4-d-2\epsilon}
\end{equation}
to generate turbulent velocity field
with Kolmogorov's scaling~\cite{DeDominicis79,Vasilyev83} (which is achieved
by choosing  $\epsilon=2$).

The stochastic Navier-Stokes equation (\ref{NS}) is a Langevin type equation leading to a MSR
operator functional which in the functional-integral form
gives rise to
the following action functional
\ba
\label{S2}
S_2&=&
{1\over 2}\int\!dt d\vx d\vx'\,\tilde{\vv}(\vx,t)
\cdot\tilde{\vv}(\vx',t)d_f(\vert\vx-\vx'\vert)\nonumber\\
&+&\int\!dt d\vx\,\tilde{\vv}\cdot\left[ -\partial_t {\bf v} -
 ({\bf v} \cdot {\nabla} ) {\bf v} + \nu_0 \nabla^2 {\bf v}\right]\,.
\ea
Combined averaging over density and velocity fluctuations yields
\be
\langle O(t)\rangle=\int\!{\cal D}[\psi,\vv]
\, O_N(1,\psi_A,1,\psi_B)\,
{\rm e}^{S_0+S_1+S_2}
\ee
for the expectation value of the observable $O$.

\section{Decay asymptotics controlled by stable fixed points}

Power counting shows that in the case in which all three reaction terms
in (\ref{S1}) have the same scaling dimension
the critical dimension of the model is two~\cite{Lee95}.
Near two dimensions, however, the drift part (\ref{S2})
of the dynamic action with the nonlocal kernel (\ref{df})
is not multiplicatively renormalizable.
Therefore
I have used the kernel function~\cite{Honkonen96}
with a local term added at the outset:
\be
\label{dT}
d_f(k)=g_{10}\nu_0^3k^{4-d-2\epsilon}+g_{20}\nu_0^3k^2\,.
\ee
Apart from rendering the field theory multiplicatively
renormalizable, the local term
also has an important physical meaning:
with a suitable choice
of the parameter $g_{20}$ it describes thermal fluctuations of the velocity
field near equilibrium.

Taking this into account, I write the renormalised action in the form
\ba
\label{fullaction}
S=&-&\int\! d\vx dt\,\Bigl\{
\psi^+_A\partial_t\psi_A+\psi^+_B\partial_t\psi_B
+\phantom{\bigl(}\!\!\!\psi^+_A\nabla (\vv\psi_A)+\psi^+_B\nabla (\vv\psi_B)
\nonumber\\
&-&\phantom{\bigl(}\!\!\!u_A\nu Z_{2A}\psi^+_A\nabla^2\psi_A-u_B\nu Z_{2B}\psi^+_B\nabla^2\psi_B\nonumber\\
&+&\lambda\nu\mu^{-2\delta}Z_4\left(\psi^+_A+\psi^+_B+\psi^+_A\psi^+_B\right)\psi_A\psi_B\nonumber\\
&-&{1\over 2}\tilde{\vv}\left[g_1\nu^3\mu^{2\epsilon}
(-\nabla^2)^{1-\delta-\epsilon}-g_2\nu^3\mu^{-2\delta}Z_3\nabla^2\right]\tilde{\vv}\\
&+&\tilde{\vv}\cdot\left[\partial_t {\bf v} +
 ({\bf v} \cdot {\nabla} ) {\bf v} - \nu Z_1\nabla^2 {\bf v}\right]\Bigr\}\nonumber\\
&+&\int\! d\vx\,\Bigl[ \kappa_A \mu^dZ_{5A}\psi^+_A+
\kappa_B \mu^dZ_{5B}\psi^+_B
-\rho \mu^dZ_{5}\psi^+_A\psi^+_B\Bigr]\,,\nonumber
\ea
in which, apart from the standard renormalisation of
the dynamic action, also the renormalisation of the initial condition - predicted by power counting
and confirmed by calculations - is introduced.

Renormalisation constants have been calculated in one-loop approximation with the
use of combined dimensional and analytic regularisation with the
parameters $\epsilon$ and $\delta=(d-2)/2$, which eventually give rise to a
two-parameter expansion of critical exponents and other physical quantities.

The unrenormalized
parameters of the initial conditions have positive canonical scaling dimensions.
This means that the corresponding running parameters grow in the long-time large-scale
limit. Therefore, some kind of partial summation of the perturbative expansion
is called for to cope with this problem. A natural way would be the use of skeleton
equations for Green functions with dressed field averages and correlation functions.
In the case of single-species annihilation reaction
$A+A\to\emptyset$ this leads to well-controlled estimates
of the behaviour of scaling functions in the long-time limit~\cite{Lee94}.
Basically, this amounts to independence of the asymptotics of the initial
density~\cite{Ohtsuki91}.
In the present case of bimolecular annihilation, however, a similar direct
summation has not been found~\cite{Lee95}. Unfortunately,
I have not been able to do any
better with an analytic solution of the set of integro-differential equations,
which can be written for the dressed
one-point and two-point Green functions of the
present model and incorporate the effect of initial conditions completely.

In the leading order in the coupling constant $\lambda$
the initial density fluctuations change the classical
rate eqs (\ref{classicrate}) by the addition of
an inhomogeneous term. This leads to the system
\ba
\label{ABrho}
{\partial_tn_A} &=&
-\lambda\nu\mu^{-2\delta}\left\{n_An_B
-{\rho\mu^d\over[4\pi\nu(u_A+u_B)t]^{d/2}}\right\}
\,,\nonumber\\
{\partial_tn_B} &=&
-\lambda\nu\mu^{-2\delta}\left\{n_An_B
-{\rho\mu^d\over[4\pi\nu(u_A+u_B)t]^{d/2}}\right\}\,,
\ea
with the initial condition:
$n_A(0)=\kappa_A\mu^dZ_{5A}$,
$n_B(0)=\kappa_B\mu^dZ_{5B}$.
For the important special case $n_A=n_B=n$ the special
Riccati's equation results. The solution is known and may
be expressed in terms
of modified Bessel functions $K_{2/(4-d)}$ and $I_{2/(4-d)}$. In this solution
the blow-up of initial conditions is controllable and the asymptotic
behaviour of the particle density may be inferred. However,
the effect of initial conditions is not fully accounted for by
the manageable system of eqs
(\ref{ABrho}), and
therefore the following results give the correct asymptotic density decay
with this provision only.

The renormalised action (\ref{fullaction}) gives rise to a system
of characteristic equations with four IR stable physical fixed
points with the following asymptotic decay of the density.

({\it i}$\,$) Gaussian fixed point
\[
g_1^*=g_2^*=\lambda^*=0\,.
\]
The Gaussian fixed point is stable, when
$
\epsilon<0
$,
$ \delta >0
$.
Asymptotic decay of the number density in terms of physical
(unrenormalised) parameters
\be
\label{G}
n(t){
{\,\,\atop
{\displaystyle\sim\atop \scriptstyle t\to\infty}}
}
{\sqrt{r_0n_0}\over [4\pi(D_{A0}+D_{B0})t]^{d/4}}
\ee
is not given by the classical solution (\ref{nn}) but is slower.
This is different from the unimolecular case~\cite{Hnatic00} in which
at the Gaussian fixed point the mean-field solution holds.
Note that there is no dependence on the rate coefficient in (\ref{G}),
but the parameter of initial correlations remains. The influence of
initial correlations becomes irrelevant and the mean-field decay $\propto 1/t$
is restored only at $d>4$~\cite{Lee95}.

({\it ii}$\,$)
Thermal fixed point
\[
g_1^*=0\,,\quad g_2^*=-32\pi\delta\,,\quad
\quad u^*={\sqrt{17}-1\over 2}\,,\quad  \lambda^*=-2\pi(\sqrt{17}-1 )\delta\,.
\]
The basin of attraction of this fixed point is
$\delta<0$, $2\epsilon+3\delta <0$.
Decay rate is faster than the initial-density-fluctuation
induced:
\be
\label{T}
n(t){
{\,\,\atop
{\displaystyle\sim\atop \scriptstyle t\to\infty}}
}
{\sqrt{r_0n_0}\over [4\pi\nu_0(\sqrt{17}-1)\tau]^{d/4}}\left({
\tau\over t}\right)^{1/2}\,.
\ee
Here, $\tau$ is a reference time scale and $\delta=d/2-1$.

({\it iii}$\,$)
Reactive kinetic fixed point
 \ba
 g_{1}^{\ast}&=&
 \frac{64\,\pi}{9}\,\frac{\epsilon\,(2\epsilon+3\delta)}
 {\epsilon+\delta},
 \quad
  g_{2}^{\ast}=\frac{64\pi}{9}\,
 \frac{\epsilon^2}{\delta+\epsilon}\,,\nonumber\\
\quad u^*&=&{\sqrt{17}-1\over 2}\,,\
\lambda^*=-{4\pi\over 3}(\sqrt{17}-1 )(\epsilon+3\delta)\,,\nonumber
\ea
is stable, when
$
\epsilon>0
$, $ -{2\over 3}\epsilon<\delta<-{1\over 3}\epsilon$.
To linear order in $\delta$, $\epsilon$,
decay exponent the same as in thermal fixed point
\be
\label{K1}
n(t){
{\,\,\atop
{\displaystyle\sim\atop \scriptstyle t\to\infty}}
}
{\sqrt{r_0n_0}\over [4\pi\nu_0(\sqrt{17}-1)\tau]^{d/4}}
\left({\tau\over t}\right)^{1/2}\,.
\ee
The independence of the exponent of time of $\delta$ and $\epsilon$
in (\ref{T}) and (\ref{K1}) is most probably an artifact of the one-loop approximation.

({\it iv}$\,$)
Passive kinetic fixed point
 \[
 g_{1}^{\ast}=
 \frac{64\,\pi}{9}\,\frac{\epsilon\,(2\epsilon+3\delta)}
 {\epsilon+\delta},
 \quad
  g_{2}^{\ast}=\frac{64\pi}{9}\,
 \frac{\epsilon^2}{\delta+\epsilon}\,,\quad
\quad u^*={\sqrt{17}-1\over 2}\,,\quad  \lambda^*=0\,,
 \]
is stable, when
$
\epsilon>0$,
$\delta>-{1\over 3}\epsilon$.
Decay rate is faster than the initial-density-fluctuation
induced here, too:
\be
\label{K2}
n(t){
{\,\,\atop
{\displaystyle\sim\atop \scriptstyle t\to\infty}}
}
{\sqrt{r_0n_0}\over [4\pi\nu_0(\sqrt{17}-1)\tau]^{d/4}}\left({
\tau\over t}\right)^{d/4(1-\epsilon/3)}\,.
\ee
Here, $\tau$ is the reference time scale. The decay exponent in (\ref{K2}) is exact.

From these results it follows that the decay exponent is a continuous
function of $\delta=d/2-1$ and $\epsilon$
 - apart from
logarithmic corrections on the basin boundaries.

As in the case of unimolecular reaction~\cite{Hnatic00},
the fixed point corresponding to the pure diffusion-limited reaction
\[
g_1^*=g_2^*=0\,,\qquad \lambda^*=-4\pi(\sqrt{17}-1 )\delta
\]
is unstable in $d<2$.
This means that any velocity fluctuations (including the ubiquitous thermal noise!) drive the system
to the advection-diffusion-controlled regime with different decay exponents.

In case of unequal initial densities the system (\ref{ABrho}) leads to a particular
form of Riccati's general equation, for which the solution seems to be unknown.
Therefore, I will not discuss this case here in the hope that a solution of this
equation in terms of known special functions may eventually be found with a reasonable
effort.

\begin{figure}[t]
\begin{center}
\epsfig{file=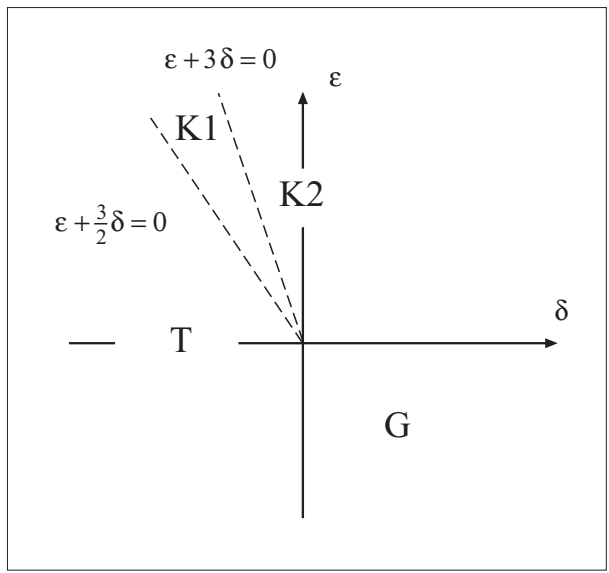,width=8cm}
      \end{center}
    \caption{Basins of attraction of the Gaussian fixed point (G), thermal fixed point (T),
the reactive kinetic fixed (K1) and the passive kinetic fixed point
(K2) in the ($\delta$, $\epsilon$) plane.}
\end{figure}

\section{Conclusions}

The main physical result of this work is that
diffusion-limited two-species
annihilation reaction is shown to be unstable to short-range (thermal)
velocity fluctuations for $d\le d_c=2$
and unstable to long-range (turbulent)
velocity fluctuations for $d\ge 2$.
Decay exponents in four stable advection-diffusion-controlled
regimes have been calculated at one-loop order.
Although the renormalisation and fixed-point analysis
are fairly straightforward, the asymptotic analysis of
scaling functions is not.
The blow-up of the effective (running)
initial conditions in the scaling functions is difficult to control,
in contrast with the single-species case, and a firm
conclusion about the asymptotics of scaling functions is not yet
available.

As to possible generalisations, it would be interesting to amend the decay analysis
by a similar treatment of a
stationary state with reactant sources.

It seems quite plausible that
the asymptotic behaviour of the density heavily depends on the localisation
of the initial density profiles; therefore, an analysis of the
problem with
localised initial conditions would be desirable.

Due to the incompressibility condition imposed on the drift
field,
the present results have a direct physical meaning at $d\ge 2$
only. Thus, the
effect of compressibility should be analysed. This, however, does not seem
to be feasible at present in the full stochastic Navier-Stokes framework.
Therefore, to make some progress in this direction, it would be interesting to analyse
the effect of velocity fluctuations with given statistics instead of dynamically
generated random drift.

%
%
%

\end{document}